# Comment on "Discreteness Effects in Lambda Cold Dark Matter Simulations: A Wavelet-Statistical View" by Romeo et al.

Adrian L. Melott
Department of Physics and Astronomy
University of Kansas
melott@ku.edu

Abstract: Essential equivalence of the conclusions of Romeo et al. with earlier work is pointed out. A possible general explanation for these conclusions is suggested.

Introduction and Conclusions

Romeo et al. (2008) performed simulations with truncated power spectra in order to investigate the effect of small—scale power in initial conditions in numerical simulations. This approach was introduced in Melott & Shandarin (1989); see also Melott & Shandarin (1990), Beacom et al. (1991), Melott & Shandarin (1993).

They conclude that discreteness effects are visible in the simulations on all scales *ε < 2d* where *ε* is the force resolution (sometimes called "softening"), and *d* is the mean interparticle separation. Although there is no sharp and sudden disappearance of discreteness effects, this conclusion is in disagreement with the usual claims of viability down to the scale *ε*. Furthermore, although not noted by Romeo et al., this is essentially the same as conclusions previously discussed by Peebles et al. (1989), Melott & Shandarin (1989), Melott (1990), Melott & Shandarin (1990), Beacom et al. (1991), Melott & Shandarin (1991), Kuhlman et al. (1996), Melott et al. (1997), and Splinter et al. (1998). Specific numerical tests of discreteness are described in the first of these, and are the primary topic of the last three; the others present visual comparisons and use truncated power spectra for comparison purposes.

The suppression of discreteness is the basis for the choice of PM codes over direct N-body and $P^3M$ methods, since it maximizes reliable resolution for given computer capacity compared with those approaches. Such suppression is particularly important where discreteness can easily spoil results (e.g. Melott 1982). This, for example, made it possible to discern for the first time filamentary superclusters in CDM (Melott et al. 1983). As noted by Romeo et al., adaptive mesh refinement may make it possible to improve upon this, but AMR needs careful, critical tests, and will still exclude initial fluctuations below the comoving initial *particle* Nyquist scale.

Discussion

A possible explanation for this limitation based on the idea of chaotic trajectories was proposed by Melott et al. (1997). Chaotic trajectories were analyzed by Suto (1991) as a function of time and the ratio *ε/d*. He found that the Lyapunov exponent $\sim \varepsilon^{-1.2}$. The criterion that uncertainties in trajectory on scale ε not grow larger than scale *d* in a Hubble Time leads to the condition *ε ≥ d*; for more details see these two papers.